\documentstyle[12pt,psfig]{article}
\newcommand{\beq}[1]{\begin{equation}\label{#1}}
\newcommand{\eeq}{\end{equation}}
\newcommand{\beqar}[1]{\begin{eqnarray}\label{#1}}
\newcommand{\eeqar}{\end{eqnarray}}
\newcommand{\lash}[1]{\not\! #1 \,}
\begin{document}
\vspace*{-2cm}
\vspace{4cm}
\begin{center}
{\bf \large Ioffe-time distribution of quarks in the photon}
\vspace{1cm}

S.~Hofmann$^1$, L.~Mankiewicz$^{2,}$ \footnote[1]
{on leave of
N. Copernicus Astronomical Center, Bartycka 18, PL--00--716 Warsaw,
Poland}
and A.~Sch\"afer$^1$
\vspace{1cm}

$^1$Institut f\"ur Theoretische Physik, J.~W.~Goethe
Universit\"at Frankfurt,
\\
Postfach~11~19~32, W-60054~Frankfurt am Main, Germany
\\
$^2$Institut f\"ur Theoretische Physik, TU-M\"unchen,
James-Franck-Stra\ss e, W-85747~Garching, Germany

\end{center}

\vspace{2cm}
\noindent{\bf Abstract:}
\vskip 0.5 cm
We have analysed the Ioffe-time distribution of quarks in virtual
photons using Operator Product Expansion of the correlation function that
determines the matrix element of the corresponding quark string operator.
The distribution for a transversally polarised photon admits a
spectral representation which can be continued to the on-shell region $p^2
=0$.
The resulting model Ioffe-time distribution turns out to be larger than
parametrisations of the available $F_2^{\gamma}$ data.
This result is linked to the slope of the quark distribution
at the origin, which is also too large.

\eject
\newpage
Many years ago Witten \cite{Witten} showed that the
Operator Product Expansion (OPE) can be applied to the description
of photon-photon scattering with the target photon on or near mass shell
while the probing photon has a large Euclidean momentum.
According to Witten's analysis the structure function
of a transversal polarised photon contains a point-like part
which grows logarithmically as a function of $Q^2$. It arises due
to the contribution of highly virtual quarks in the quark loop
and can be fully predicted in perturbative QCD.
The contribution of low-virtuality quarks describes
the interaction of the probe with the hadronic component of
the photon, which is subdominant at asymptptically large $Q^2$.

It is necessary to understand this hardronic contribution
to the photon structure function better because all experiments probe
the photon structure at $Q^2$ values, for which its hadronic part
is non-neglible. Because it has not yet been possible
to calculate it from first principles, Vector Meson Dominance
(VDM) type models are used to estimate its magnitude.
To link photon properties more directly to QCD, Gorsky et al. \cite{Gorsky}
calculated the hadronic part of the photon structure function applying OPE
to the photon-photon forward Compton amplitude, which resulted in an expansion
in inverse powers of the target photon virtuality.
This expansion admits a spectral representation which allowed an analytic
continuation to the quasi real target photon regime $p^2 \sim 0$.

However, the problem with parton distributions in momnetum space is that
for a given value of the momentum fraction $u$
they get contributions from both small and large longitudinal distances.
To circumvent this problem the authors of \cite{Iof0,Tera}
proposed an alternative representation, which was already applied to the
quark distribution of the nucleon \cite{BGM}.
It turned out that in complete duality to twist-2 parton distributions
in momentu space, twist-2 parton distributions in coordinate space
can be introduced. The coordinate space variable, called Ioffe-time,
measures the longitudinal light-cone distance between the points
where the hard probe was absorbed respectively emitted by the target.
Derivatives of Ioffe-time distributions at the origin are given by moments
of some coresponding structure functions, or equivalently \cite{Collins,bal88}
by matrix elements of twist-2 operators of increasing dimension. From
the theoretical point of view, this connection makes Ioffe-time distributions
much easier to analyse than parton distributions in momentum space.
%
%

In the present paper we concentrate on quark
distributions $q^\lambda(u)$ and ${ \bar q}^\lambda(u)$ in a photon
with momentum $p$ and polarization $\lambda$.  Following reference
\cite{BGM} we introduce the Ioffe-time distribution of quarks inside
the photon $Q^\lambda(z)$, which is given by the photon matrix element
of the quark string operator ${\hat O}(\Delta) = {\bar \Psi}(\Delta)
{\not\!  n \,}[\Delta;- \Delta] \Psi(-\Delta)$,
\begin{eqnarray}
2 i (p \cdot n) Q^\lambda(z)
& = &  \langle p \lambda \mid {\hat O}(\Delta)\mid p \lambda \rangle
\nonumber \\
& = & 2 (p \cdot n) \int_0^1 du\,\Big[q^\lambda (u) \, \exp{(iuz)} -
{\bar q}^\lambda (u) \, \exp{(-iu z)}\Big] \, ,  \nonumber \\
\label{One}
\end{eqnarray}
where for notational convenience we have introduced two parallel
light-like vectors $\Delta_\mu$ and $n_\mu$ satisfying
$\Delta^2 = n^2 = n \cdot \Delta = 0$. Our convention is such that for
any vector $a$, $n \cdot a = a^+$.
In equation (\ref{One})
$z=2 \, p \cdot \Delta$ and $[\Delta;- \Delta]$ denotes
the path-ordered exponential which ensures gauge invariance of the
quark distribution.
As we have to respect both QED and QCD \cite{HoMaSch} gauge symmetry
$[\Delta;- \Delta]$ becomes the product of colour and electromagnetic parts
\begin{equation}
[ \Delta;- \Delta] =
{\large P}\exp {\left[i g \int_{-1}^1 d\xi \Delta \cdot
A(\xi\Delta)\right]} \, \cdot \,
{\large P}\exp {\left[-i e \int_{-1}^1 d\eta \Delta \cdot
{\tilde A}(\eta\Delta)\right]} \, ,
\label{PQCD}
\end{equation}
where $A_\mu(x)$ is the gluon field and
${\tilde A}_\mu(x)$ denotes the photon field.
In the following, however, we will adopt the Schwinger gauge
$x_\mu {\tilde A}_\mu(x) = 0$ for the electromagnetic field, such that the
second factor becomes one, and never explicitely enters the calculations.

A Taylor expansion of ${\hat O}(\Delta)$ in $\Delta_{\mu}$
results in local operators of twist 2. The first term in this expansion
is given by the traceless part of the quark energy-momentum tensor
and the matrix element of its QCD part describes the longitudinal momentum
fraction carried by quarks in the target.
The normalization of Ioffe-time distributions is such that
this matrix element is given just by the derivative of $Q^\lambda(z)$ at $z =
0$.
As the photon is even under charge conjugation (\ref{One}) becomes
\begin{equation}
\langle p \lambda \mid {\hat O}(\Delta)\mid p \lambda \rangle
= 2 i (p \cdot n) \int_0^1 du\,\Big[q^\lambda (u) + {\bar q}^\lambda (u)
\Big] \sin(uz)
\, ,
\label{Two}
\end{equation}
which clearly shows that the Ioffe-time $z$ is the Fourier-conjugate to the
usual Bjorken variable $u$.

Using the LSZ reduction formula we calculate the photon matrix
element (\ref{One})
as a three-point correlation between two electromagnetic currents
and the operator ${\hat O}(y;\Delta)$ centered at a point y:
\begin{equation}
{\hat O}(y;\Delta) =
{\bar \Psi}(y+\Delta) {\not\! n \,}[y+\Delta;y- \Delta] \Psi(y-\Delta)
\, .
\end{equation}
Denoting by $M_{\mu\nu}(p,k)$ the three-point correlation function
\begin{equation}
M_{\mu,\nu}(p,k) = i^2e^2
\int d^4x \, d^4y \, \exp{(ipx + iky)}
\langle 0|T(j_{\mu}(x){\hat O}(y; \Delta)j_{\nu}(0))|0 \rangle \, ,
\label{Mmn}
\end{equation}
we  obtain
\begin{equation}
\langle p \lambda \mid {\hat O}(y;\Delta)\mid p \lambda \rangle =
\epsilon_{\mu}^*(p,\lambda) M_{\mu\nu}(p,k=0)
\epsilon_{\nu}(p,\lambda)
\label{MatEl}
\end{equation}
We remind the reader that (\ref{MatEl}) holds only in the Schwinger
gauge for the photon field; in general some extra terms arise.

With the Ward identity derived in the
appendix of \cite{BGM} we can write $M_{\mu \nu}(p,k)$ as
a sum involving two- and three point correlators:
\begin{eqnarray}
&& M_{\mu\nu}(p,k) =  \nonumber \\
&&{}e^2 \int d^4x\, \exp{(i p \cdot x)} \int d^4y\,
\exp{(i k \cdot y)}
\,\frac{n \cdot x}{k \cdot x} \int_{-1}^1 d\xi\ \langle 0 \mid
T\Big\{ j_{\mu}(x) {\bar \Psi}(y+ \Delta)
\nonumber\\
&& {}\times [y+ \Delta;y+ \xi \Delta]
 g \gamma^\rho \Delta^\sigma G_{\rho\sigma}
(y+ \xi \Delta)
[y+ \xi \Delta; y- \Delta]
 \Psi (y- \Delta) j_{\nu}(0)\Big\} \mid
0 \rangle  \nonumber \\
&&{}+ i e^2 \int d^4x\, \exp{(i p \cdot x)}
 \frac{n \cdot x}{k \cdot x}
\Big[ \exp{(i k \cdot x)}
\langle 0 \mid T (j_{\mu}(x; x- 2 \Delta) j_{\nu}(0))
\mid 0 \rangle \nonumber \\
& &{}-  \langle 0 \mid T (j_{\mu}(x) j_{\nu}( 2 \Delta; 0))
\mid 0 \rangle + {}(\Delta \to - \Delta) \Big]\, ,
\nonumber \\
\label{W4}
\end{eqnarray}
where $j_\mu(x,y)$ is the point-splitted electromagnetic current
operator,
\begin{equation}
j_\mu(x,y) = {\bar \Psi}(x)\gamma_\mu [x;y] \Psi(y) \, .
\end{equation}
Note that the momentum $k_{\mu}$ appears in the denominators in (\ref{W4}) and
therefore it can be taken to zero only at the end of calculation after
all potentially divergent terms canceled. Expression
(\ref{W4}) is formally equivalent to (\ref{Mmn}) but allows
to apply directly the techniques developed previoulsy in \cite{BGM}.

In the following we distinguish between Ioffe-time distributions in
transversally (T) and longitudinally (L) polarised photons,
defined by
\begin{eqnarray}
\frac{1}{2} \sum_{\lambda=1,2}
\langle p \lambda \mid {\hat O}(\Delta)\mid p \lambda \rangle
& = & 2 i (p \cdot n) Q^T(z) \nonumber \\
\langle p L \mid {\hat O}(\Delta)\mid p L \rangle
& = & 2 i (p \cdot n) Q^L(z)
\label{Def}
\end{eqnarray}
We introduce the vector
\begin{equation}
{\tilde p}_\mu = p_\mu - \frac{1}{2} \frac{p^2}{p\cdot n} n_\mu \, ,
\end{equation}
such that ${\tilde p}^2 = 0$. The polarisation tensor can
than be written in terms of transverse and
longitudinal parts, $d_{\mu\nu} = d_{\mu\nu}^T - d_{\mu\nu}^L$:
\begin{eqnarray}
d_{\mu\nu}^T & = & \sum_{\lambda=1,2}\epsilon_\mu^*(p,\lambda)
\epsilon_\nu(p,\lambda) = - g_{\mu\nu} +
\frac{n_\mu {\tilde p}_\nu + n_\nu {\tilde p}_\mu}{p\cdot n} \nonumber \\
d_{\mu\nu}^L & = & - \epsilon_\mu^*(p,L)\epsilon_\nu(p,L) =
-p^2 \frac{n_\mu n_\nu}{(p.n)^2} \, .
\label{Polsum}
\end{eqnarray}

Our main task is using OPE to calculate the three-point correlation function
(\ref{Mmn}) for the case of massless quarks
up to the leading non-perturbative corrections of dimension 4,
\begin{equation}
Q^{T,L}(z) = Q^{T,L}_0(z) + \frac{1}{(-p^2)^2} Q^{T,L}_4(z) \,
\label{QOPE}
\end{equation}
where $(-p^2)$ is the virtuality of the target photon and $\lambda$ its
polarisation.

We shall first consider the situation for transverse polarisation.
The perturbative contribution $Q^T_0(z)$ to the matrix element
(\ref{Def}) arises from the last two terms in (\ref{W4}).
The explicit calculation \cite{HoMaSch} gives:
\begin{equation}
Q^T_0(z)
= \frac{3 \alpha}{ \pi} \, \sum_q e_q^4 \, \int du \, \sin(u \cdot z)
\, \left[(u^2 + {\bar u}^2) \, \log{(\frac{\mu^2}{-p^2})} - (u^2 +
{\bar u}^2) \log{(u {\bar u})} - 1 \right] \, ,
\label{QT0}
\end{equation}
where $\mu^2$ is the ultraviolet cut-off corresponding to the
normalisation point of the operator ${\hat O}$, and $\alpha =
\frac{1}{137}$ is the elctromagnetic coupling constant. Assuming a relatively
low normalisation point $\mu^2 \sim$ a few GeV$^2$, the sum over active quark
flavours runs over u,d and s quarks and $\sum e_q^4 = \frac{2}{9}$.  Note
that (\ref{QT0}) has been obtained using dimensional regularisation
and minimal subtraction of the ultraviolet divergence. As it is well
known, only the part proportional to $\log{(\frac{\mu^2}{-p^2})}$ does
not depend on the renormalisation scheme.

Let us now consider the $Q_4^T(z)$ term in OPE (\ref{QOPE}).
In general $Q_4^T(z)$ will contain terms proportional to
the VEV of the dimension-4 operator $\langle \frac{ \alpha_s}{ \pi} G^2
\rangle$
and terms involving dimension-4 correlators which may arise from
bilocal power corrections. BPC's are taken into account to avoid
ill-defined infrared singularities.
Following \cite{Kolesnichenko} this
can be done by introducing the ``effective propagator''
in the form:
\begin{eqnarray}
S^{ab}_{ij}(k) & = & \frac{ \delta^{ab}}{3 \cdot 36}
\, \langle \frac{ \alpha_s}{ \pi} G^2 \rangle
\, (k \cdot x) \Big[{ \lash \Delta}_{ij}
\, [ \log{( \frac{s_R(-x^2)}{4})} - \frac{5}{3} + 2 \gamma_E]
\nonumber \\ & + &
\frac{1}{2}(\Delta \cdot x) \frac{\lash{x}}{(-x^2)} \Big]
- \frac{ \delta^{ab}}{3} \, (k \cdot x) { \lash \Delta}_{ij}
\, [f_R^2 m_R^2 - \frac{ \alpha_s s_R^2}{480 \pi^3}] + \dots \, ,
\nonumber \\
\label{sef1}
\end{eqnarray}
where eclipses stand for terms involving higher powers of $(k\cdot x)$
and $k^2$, and $m^2_R = 1$ GeV$^2$, $s_R = 1.4$ GeV$^2$ are parameters
found in \cite{Kolesnichenko}.
This leads to the coefficient function $Q_4^{T}(z)$
\begin{eqnarray}
Q_4^T(z)
& = &
- \frac{4 \pi \alpha}{144}
\, \langle \frac{ \alpha_s}{ \pi} G^2 \rangle
\, \sum_q e_q^4
\, z \, \int_0^1 du \, \cos(u \cdot z)
\, \Big [5 \delta(u) - \delta({ \bar u}) + 16[ \frac{1}{u}]_+ -4
\Big ]
\nonumber \\
& + &
4 \pi \alpha \, \sum_q e_q^4 \,  z
\Big [ \frac{1}{9} \, \langle \frac{ \alpha_s}{ \pi} G^2 \rangle
\, \Big ( \log{( \frac{s_R}{-p^2})} + 2\gamma_E - \frac{5}{3} \Big)
+ \frac{ \alpha_s s_R^2}{120 \pi^3} - 4 f_R^2 \, m_R^2 \Big ]\, ,
\nonumber \\
\label{MGTfull}
\end{eqnarray}
where for any function f(u)
\begin{equation}
\int_0^1 du \left[ \frac{1}{u}\right]_+f(u) \equiv
\int_0^1 du\, \frac{1}{u} (f(u) - f(0)) \, .
\label{plus}
\end{equation}
In the case of a logitudinally polarized target photon
the calculation is much less complicated because infrared divergences
are absent. The dimension zero coefficient  $Q_O^L(z)$ is given by
\begin{equation}
Q_0^L(z)
=
\frac{12 \alpha}{ \pi} \, \sum_q e_q^4 \,
\int_0^1 du \, \sin(u \cdot z) \, u \, { \bar u} \, ,
\end{equation}
while for the dimension four coefficient $Q_4^L(z)$ we obtain [??]
\begin{equation}
Q_4^L(z)
=
\frac{4 \pi \alpha}{18}
\, \langle \frac{ \alpha_s}{ \pi} G^2  \rangle
\, \sum_q e_q^4
\, z \, \int_0^1 du \, \cos(u \cdot z) \, .
\end{equation}

At this point it is possible to compare our exact results with that of
\cite{Gorsky}, where the photon structure function $F_2(u)$
was only calculated in the region of intermediate $u$.
Within the logic of the present calculation $F_2(u)$
can be obtained from the expression for Ioffe-time distribution by extracting
the integrand in (\ref{One}) and expanding it to the first order in $z$.
As can be easily seen our dimension zero coefficients $Q_0^T(z)$ and
$Q_0^L(z)$
are consistent with the results of ref.\cite{Gorsky}.
For comparison of the dimension four coefficients
we have to neglect all singular contributions to $F_2(u)$ concentrated
at the boundary values $u=0$ and $u=1$.
In general such singular contributions are
crucial in calculations of moments of the structure function
\cite{Kolesnichenko,BelBlok}. Certainly, in the case of
Ioffe-time distributions they cannot be neglected.
In ref.\cite{Gorsky} BPC's were not taken into account.
However, the dimension four contribution can be reproduced
if the $[\frac{1}{u}]_+$ distribution in (\ref{MGTfull}) is interpreted
simply as function $\frac{1}{u}$.

To continue our predictions for transversally polarised on-shell photons
we construct a Mandelstam dispersion representation in $(-p^2)$:
\begin{equation}
Q^T(z) =
A_0(z) + \int_0^\infty ds \frac{A_1(s;z)}{s-p^2}
\, + \int_0^\infty ds_1 \int_0^\infty ds_2\,
\frac{A_2(s_1,s_2;z)}{(s_1-p^2)(s_2-p^2)}
\, .
\label{disp3}
\end{equation}
In the region $s_R < -p^2 \ll \mu^2$, spectral densities $A_1(z;s)$,
$A_2(s_1,s_2,z)$ can be found through discontinuities, in
$p_1^2 = p^2$ and $p_2^2 = (p+k)^2$, of loop diagrams
which contribute to $M_{\mu,\nu}(p,k)$, equation (\ref{Mmn}).
We make the standard approximation to represent $A_0$, $A_1$ and
$A_2$ by a contribution from the $\rho$ meson plus a continuum
contribution with a treshold $s_0$. In this way we obtain
\begin{eqnarray}
A_0(z) & = & C_0(z) + C_1(z) \nonumber \\
A_1(s;z) & = & 0 \nonumber \\
A_2(s_1,s_2;z) & = & \delta(s_1-m^2_\rho)\delta(s_2-m^2_\rho)\left(C_2(z) +
\frac{1}{2}s_0^2 C_1(z) \right)\nonumber \\
& + &
\delta(s_1-s_2)\Theta(\mu^2-s_1)\Theta(s_1-s_0)\sqrt{s_1s_2} C_1(z)\nonumber
\\
& - & C_3(z) \delta(s_1-m^2_\rho) \Theta(\mu^2-s_2)\Theta(s_2-
s_R)\frac{1}{s_2} \, ,
\label{disp5}
\end{eqnarray}
with the coefficient functions
\begin{eqnarray}
C_0(z)
& = &
-\frac{3 \alpha}{ \pi} \, \sum_q e_q^4
\,\int_0^1 du \, \sin(u \cdot z) \, [(u^2+{ \bar u}^2)
\, \log{(u \, { \bar u})} +1]
\nonumber \\
C_1(z)
& = &
\frac{3 \alpha}{ \pi} \, \sum_q e_q^4
\, \int_0^1 du \, \sin(u \cdot z) \, (u^2+{ \bar u}^2)
\nonumber \\
C_2(z)
& = &
- \frac{4 \pi \alpha}{144}
\, \langle \frac{ \alpha_s}{ \pi} G^2 \rangle
\, \sum_q e_q^4
\, z \, \int_0^1 du \, \cos(u \cdot z)
\, \Big [5 \delta(u) - \delta({ \bar u}) + 16[ \frac{1}{u}]_+ -4
\Big ]
\nonumber \\
& + &
4 \pi \alpha \, \sum_q e_q^4 \,  z
\Big [ \frac{1}{9} \, \langle \frac{ \alpha_s}{ \pi} G^2 \rangle
\, ( 2\gamma_E - \frac{5}{3} )
+ \frac{ \alpha_s s_R^2}{120 \pi^3} - 4 f_R^2 \, m_R^2 \Big ]
\nonumber \\
C_3(z)
& = &
4 \pi \alpha \, \langle \frac{ \alpha_s}{ \pi} G^2 \rangle
\, \sum_q e_q^4 \, \frac{1}{9} \, z \, .
\label{disp2}
\end{eqnarray}
This provides a representation of $Q^T(z,-p^2)$ valid for all $-p^2$,
\begin{eqnarray}
Q^T(z,-p^2)
& = &
\frac{ C_2(z)
+ \frac{1}{2} \, s_0^2 \, C_1(z)}{(m_\rho^2-p^2)^2}
\nonumber \\
& + &
C_0(z) + C_1(z)
+ \Big [ \log{( \frac{\mu^2}{s_0-p^2})} + \frac{p^2}{s_0-p^2} \Big]
\, C_1(z)
\nonumber \\
& - &
\frac{1}
{(-p^2) \, (m_\rho^2-p^2)}
\, \log{( \frac{s_R-p^2}{s_R})}
\, C_3(z) \, ,
\label{disp6}
\end{eqnarray}
which can be used to extrapolate to $p^2=0$.
For numerical calculations we chose the normalisation scale $\mu^2 = 4$
GeV$^2$, and the other parameters equal to their standard values $s_0 =
1.5$ GeV$^2$, $s_R = 1.4$ GeV$^2$, $m^2_R = 1$ GeV$^2$, $\langle
\frac{\alpha}{\pi} G^2\rangle = 0.012$ GeV$^4$.

Note that the
representation (\ref{disp6}) has a clear physical interpretation. The first
two lines arise from diagonal $\rho$ meson to $\rho$ meson and
continuum to continuum transitions. The contribution in the third line is
due to an off-diagonal $\rho \rightarrow$ continuum transition.
The $\rho \rightarrow \rho$ transition is responsible for only
half of the distribution at small z, the rest coming from
continuum $\rightarrow$ continuum and, much
smaller, non-diagonal transitions.
For small z, in the region where
the present calculation can be trusted, the model Ioffe-time
distribution follows practically a straight line i.e., it is
practically determined in this region by the first derivative at $z=0$
which equals to the momentum fraction carried by quarks.

The next step is to supplement the model Ioffe-time distribution
for an on-shell photon by a pointlike contribution according to
Witten's analysis, see \cite{Witten}.
Within the present formalism it amounts to the replacement of the
factorisation
scale $\mu^2$ by the virtuality of the hard photon $Q^2$, and adjusting the
$-p^2$ independent term $C_0(z)$ accordingly
\begin{equation}
C_0(z) = \frac{3\alpha}{\pi} \, \sum_q e_q^4 \int_0^1 du \, \sin(u \cdot z) \,
u \, [(u^2 + {\bar u}^2) \, \log(\frac{1}{u^2}) \, + 8u{\bar u} -2]
\label{C0}
\end{equation}
which reproduces exactly the imaginary part of the box graph
\cite{Gorsky}.
Furthermore, for comparison with relatively low
$Q^2 = 5.2$ GeV$^2$ data it is appropriate to neglect the perturbative
evolution
and treat the
resulting Ioffe-time distribution as a low-scale, non-perturbative input.
As can be seen from Figure 4
the structure function of the real photon turns out to be larger
in the region of small z than
parametrisations \cite{Schuler} of the available $F_2^{\gamma}$ data.

As a consequence the momentum fraction carried
by quarks inside the transversal polarised $\rho$ meson comes
out too large as well. The Ioffe-time distribution $Q^{\rho T}(z)$
for quarks in the transversal polarised $\rho$ meson is related to the
double $\rho$ meson pole, renomalised by the coupling constant of the
$\gamma \rightarrow \rho$ transition
\begin{equation}
C_2(z) + \frac{1}{2} \, s_0^2 \, C_1(z) = \sum_q e^4_q \,
\frac{4 \pi \alpha}{g^2_V} \, m_\rho^4 \, Q^{\rho \, T}(z) \, ,
\label{Qrho}
\end{equation}
where $\frac{g^2_V}{4 \pi} \sim 1.27$.  The model thus predicts that
essentially all longitudinal momentum is carried by quarks.  Similiar problems
are known from QCD sum rules calculations of the magnitude of longitudinal
momentum fraction carried by quarks in a hadronic target
\cite{Kolesnichenko,BelBlok}, although the recent calculation of the gluonic
momentum fraction gave a reasonable result \cite{BGMS}.  Moreover, due to
universality of OPE, in each case the problem can be traced down to the
apparently too small magnitude of negative bilocal correction (\ref{sef1}).
Clearly, we have ecountered the same situation in the present context, where
it
can be equivalently related to a too small negative contribution from the
non-diagonal transitions. When we ad hoc renormalize our Ioffe-time
distribution by a factor 0.6 - 0.7, the plausible value of the momentum
fraction carried by quarks in hadronic targets, the result agrees nicely with
the $F_2^{\gamma}$ data, leaving still some room for sea quark contributions.
Of course such a step cannot be justified within the present formalism.

{\bf Acknowledgments}
This work was supported in part by BMBF, DFG., and by KBN grant 2~P302~143~06.
A.S. acknowledges support from the MPI f\"ur Kernphysik, Heidelberg.  One of
us
(LM) is grateful to Maria Krawczyk and Vladimir Braun for iluminating
discussions.

\clearpage

\section*{Figure captions}
\begin{description}
\item[Fig.~1]
Real photon
structure function $\frac{1}{\alpha} F_2^{\gamma}(z)$ (thick line)
as a function of the Ioffe-time z at $Q^2=5.2$ GeV$^2$ in comparison
with a fit \cite{Schuler} of the available $F_2^{\gamma}$ data (dashed line).
The dotted lines indicate the uncertainty of the $F_2^{\gamma}$ data.
\label{Fres2}
\end{description}

\clearpage
\begin{figure}[p]
\centerline{\psfig{figure=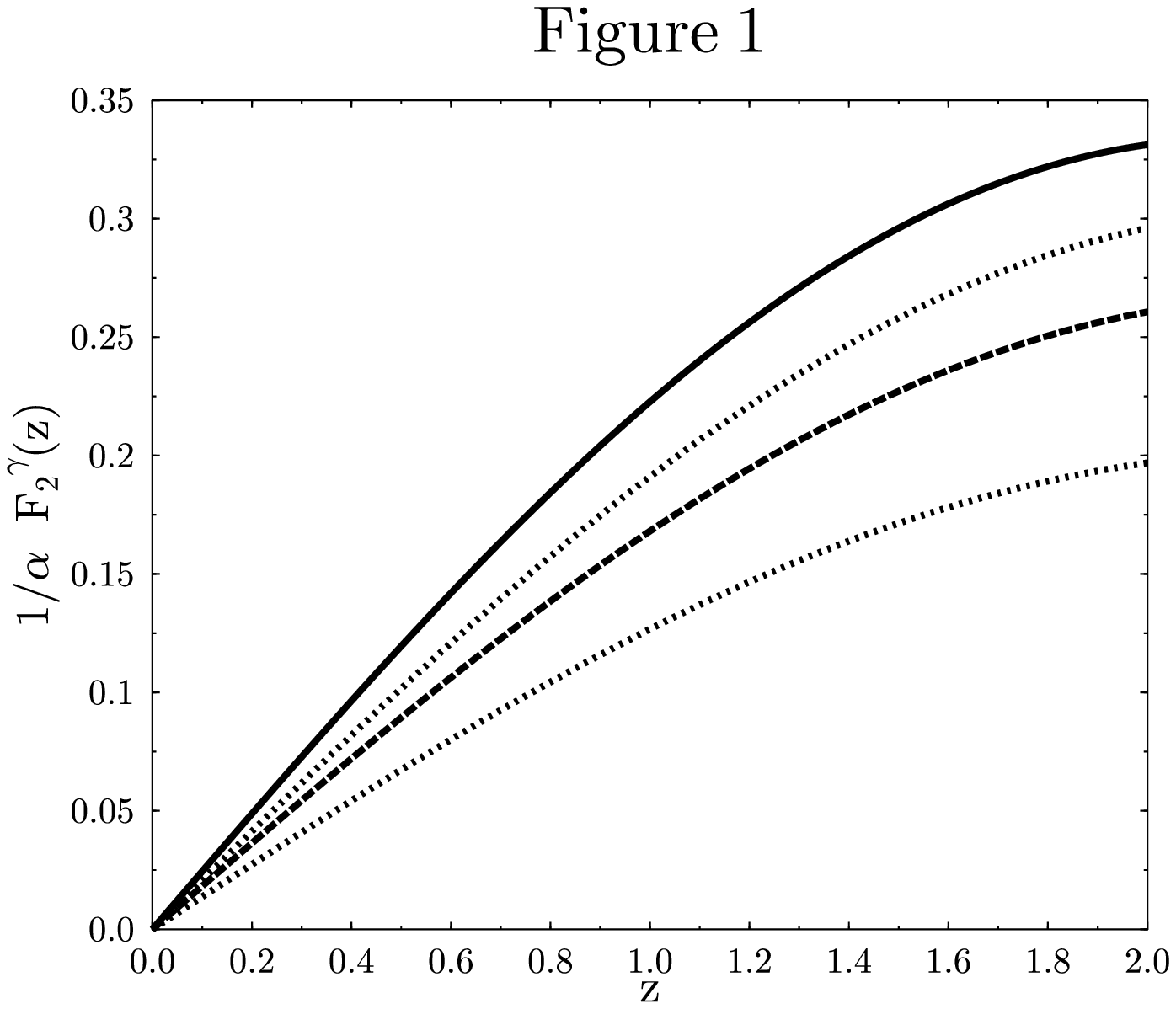,height=15cm}}
\end{figure}

\end{document}